\documentclass[preprint,prb,amsmath,amssymb,showkeys]{revtex4}

\usepackage{graphicx}
\usepackage{threeparttable}
\usepackage{array}
\usepackage{amsfonts}
\usepackage{amssymb}
\usepackage{mathtext}
\usepackage{amsmath}
\usepackage{multirow}
\usepackage{rotating}
\usepackage{ulem}

\begin{document}



\title{Protein folding: Complex potential for the driving force in
a two-dimensional space of collective variables}



\author{Sergei F. Chekmarev}

\affiliation {Institute of Thermophysics, 630090 Novosibirsk, Russia and \\Department of
Physics, Novosibirsk State University, 630090 Novosibirsk, Russia}



\date{\today}

\begin{abstract}
Using the Helmholtz decomposition of the vector field of folding fluxes in a
two-dimensional space of collective variables, a potential of the driving force for
protein folding is introduced. The potential has two components. One component is
responsible for the source and sink of the folding flows, which represent, respectively,
the unfolded states and the native state of the protein, and the other, which accounts
for the flow vorticity inherently generated at the periphery of the flow field, is
responsible for the canalization of the flow between the source and sink. The theoretical
consideration is illustrated by calculations for a model $\beta$-hairpin protein.

\end{abstract}

\maketitle


\section{Introduction}
Protein folding is a reaction in which a protein attains its functional (native) state;
the unfolded state of the protein plays a role of the reactant, and the native state
plays a role of the product. In contrast to simple chemical reactions, where the
reactant, transition state and product are well defined species, the folding reaction is
characterized by ensembles of unfolded and transition states. Accordingly, folding
pathways are numerous and diverse. The understanding of this difference between the
protein folding and chemical reactions has led to a statistical view of protein folding
\cite{BryngelsonWolynes87,ShakhnovichFarztdinovGutinKarplus91,LeopoldMontalOnuchic92,
BryngelsonOnuchicSocciWolynes95,DillChan97,Shakhnovich97,OnuchicLuthey_SchultenWolynes97,
DobsonSaliKarplus98,DSSDK00,SheaBrooks01,SchulerEverettLipmanEaton02,Gruebele2002,
MelloBarrick04,Karplus11} (for review, see Refs.
\onlinecite{OnuchicLuthey_SchultenWolynes97, DobsonSaliKarplus98,DSSDK00,SheaBrooks01}
and \onlinecite{DillOzkanShellWeikl08,DillMacCallum12, FinkelsteinPtitsyn02}). According
to this view, the folding process is governed by the interplay of the protein potential
energy and conformation entropy, forming a "funnel-like" energy surface. The energy and
entropy both decrease from the unfolded states of the protein to its native state but in
different manner, so that a free energy barrier is formed that separates the native-like
states from the unfolded states.

A tempting and widely used approach to study the folding reaction is to construct a free
energy profile as a function of a single reaction coordinate
\cite{BerezhkovskiiSzabo05,RheePande05,BestHummer2006,KrivovKarplus06,BeckDaggett07,
ChoderaPande11,PetersBolhuisMullenShea13}. However, although the diversity of folding
pathways is taken into account (as the entropy part of the free energy), this approach is
generally limited to the case of a single reaction channel, when the folding pathways can
be organized in a "transition tube" \cite{E_Vanden-Eijnden10}. A more powerful method to
characterize the diversity of the pathways is the construction of the free energy surface
(FES) of the folding reaction as a function of two collective variables
\cite{SheaBrooks01,SocciOnuchicWolynes98,BoczkoBrooks95,ChekmarevKrivovKarplus05}. One
variable is usually chosen to describe the protein compaction during folding (e.g., the
radius of gyration, $r_{\mathrm {g}}$), and the other its proximity to the native
conformation (e.g., the fraction of the native contacts, $f_{\mathrm {nat}}$). To
construct the FES, the probability for the system to be at a current point of the
conformation space $P(f_{\mathrm {nat}},r_{\mathrm {g}})$ is calculated and then, using
the Boltzmann hypothesis, is converted into the free energy
\begin{equation} \label{eq0}%
F(f_{\mathrm {nat}},r_{\mathrm {g}})= %
-k_{\mathrm{B}}T \ln P(f_{\mathrm {nat}},r_{\mathrm {g}})
\end{equation}
where $T$ is the temperature, and $k_{\mathrm{B}}$ is the Boltzmann constant. A
shortcoming of the FES thus determined is that it does not show the direction of the
motion, i.e., the protein can have the same probability to be at some point of the
conformation space when it goes towards the native state or requires partial unfolding to
reach the native state. To gain a closer insight into folding dynamics, we have recently
introduced a "hydrodynamic" description of the folding process
\cite{ChekmarevPalyanovKarplus08}. In this approach, similar to the FESs, the folding
process is considered in a reduced space of collective variables, and, similar to Markov
state models (MSMs) \cite{Schutte99,SwopePiteraSuits04,WeberPande11}, disconnectivity
graphs (DGs) \cite{BeckerKarplus97,KrivovKarplus04,Wales03} and protein folding networks
(PFNs) \cite{RaoCaflisch04,Noe-Weikl,BowmanVoelzPande11}, the calculated folding
trajectories are used to determine probabilities of transitions between the protein
states. On the one hand, in comparison to the MSMs, this allows consideration of more
complex (non-Markovian) kinetics, and on the other, in comparison to the DGs and PFNs, to
arrange the protein states according to their distribution in the conformation space
\cite{KalginCaflischChekmarevKarplus13}. The probabilities of transitions are organized
as local flows between the points of the (reduced) conformation space. Given the flows,
it is possible to construct the vector field and streamlines of the folding flows,
similar to how it is done in hydrodynamics \cite{LandauLifshitz87}. Then, the process of
protein folding can be viewed as a motion of a folding "fluid", with the density of the
fluid being proportional to the probability for the system to be at the current point of
the conformation space. In equilibrium conditions, the local flows of transitions become
small, or vanish, due to detailed balance \cite{KalginCaflischChekmarevKarplus13}.
Therefore, the hydrodynamic description is most efficient for nonequilibrium conditions,
particularly if detailed balance is absent on the overall scale. Such is the case when
the native state is essentially stable; then the unfolding events are rare, and the
process of folding is reduced to the first-passage folding. Correspondingly, the folding
reaction is represented by a steady flow of the folding fluid from the unfolded states of
the protein to its native state. For the equilibrium conditions, when the protein
repeatedly folds and unfolds, the segments of the equilibrium trajectory between these
states can be selected for the first-passage paths from the unfolded to native state, as,
for example, in the transition-path theory (TPT)
\cite{E_Vanden-Eijnden06,E_Vanden-Eijnden10}.

The hydrodynamic approach has been successfully applied to the study of folding dynamics
of several model proteins - an $\alpha$-helical hairpin (a lattice model)
\cite{ChekmarevPalyanovKarplus08}, a SH3 domain (a C$_{\alpha}$ model)
\cite{KalginKarplusChekmarev09,KalginChekmarev11}, and beta3s-miniprotein (all-atom
simulations) \cite{KalginCaflischChekmarevKarplus13}. It has been found that although the
general behavior of the folding flow is consistent with the FES landscape, i.e., the flow
is directed from the unfolded states to the native state and mostly concentrated at the
bottom of the valley that connects these states, the distribution of the local flows is
very different from what the FES suggests
\cite{ChekmarevPalyanovKarplus08,KalginKarplusChekmarev09}. Moreover, local flow vortices
can be formed that do not necessarily leave fingerprints on the FES
\cite{ChekmarevPalyanovKarplus08,KalginKarplusChekmarev09,KalginChekmarev11,
KalginCaflischChekmarevKarplus13}, and such vortical flows can have many properties of
turbulent flows of a fluid \cite{KalginChekmarev11}.  In other words, the FES does not
present a true potential for folding flows.

The knowledge of the local flows offers a possibility to determine such a potential on
the basis of the Helmholtz decomposition \cite{ArfkenWeber95} of the vector field of the
flows. In the present paper, considering the first-passage folding of a model
$\beta$-hairpin protein taken as an example, we show that this potential intrinsically
has two components. One component is responsible for the source and sink of the folding
flow and the other for the canalization of the flow between the source and sink.

The paper is organized as follows. Section 2 describes the protein model, the simulation
method, the hydrodynamic approach, and the choice of collective variables. Also, it
presents some results of the simulations to show that the present protein model leads to
a representative picture of protein folding (the FES as a function of $f_{\mathrm {nat}}$
and $r_{\mathrm {g}}$, the mean first-passage time as a function of temperature, and the
distribution of the first-passage times). Section 3 presents the main results of the work
and their discussion:  the FES and the vector flow field in the collective variables
(3.1), the potential functions for folding fluxes (3.2), and the interpretation of these
functions (3.3). Section 4 summarizes the results of the work.

\section{Protein Model, Simulation Method and Preliminary Results}
\begin{figure}[h]\centering%
\resizebox{0.3\linewidth}{!}{ \includegraphics*{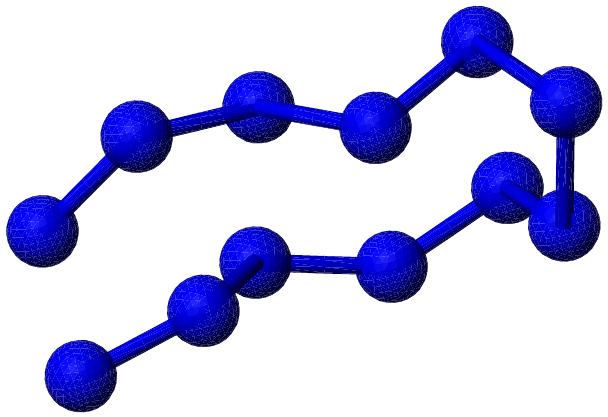}}%
\caption{The "bead" model of the native conformation of the 2evq {\it de novo} protein
\cite{2evq}.} \label{native}
\end{figure}

Determining the potential with the Helmholtz decomposition requires a smooth vector field
of folding flows and, accordingly, a large number of folding trajectories to be
simulated. To perform simulations at a reasonable cost, we considered a short
fast-folding protein in the framework of a minimalist model. The goal was to have a
representative picture of the folding dynamics rather than to describe folding of a
particular protein. Specifically, a twelve-residue $\beta$-hairpin protein 2evq
\cite{2evq} (Fig. \ref{native}) was chosen as a model system, and the interactions
between the residues were characterized using a C$_{\alpha}$-based G\={o}-like model,
i.e., the residues were represented by monomers (beads) centered at the C$_{\alpha}$
atoms, with the interaction between the monomers determined by the structure of the
native state of the protein \cite{Go83}. Specifically, the interaction potential of Ref.
\onlinecite{HoangCieplak00} was used
\begin{eqnarray} \nonumber
U=\sum_{1}^{N-1}[k_{1}(r_{i,i+1}-d_{0})^{2}+k_{2}(r_{i,i+1}-d_{0})^{4}]\\%
+\sum_{i+1<j}^{\mathrm{NAT}} 4 \varepsilon
\left[\left(\frac{\sigma_{ij}}{r_{ij}}\right)^{12}-\left(\frac{\sigma_{ij}}{r_{ij}}\right)^{6}
\right] \nonumber \\%
+\sum_{i+1<j}^{\mathrm{NON}} 4 \varepsilon
\left[\left(\frac{\sigma_{0}}{r_{ij}}\right)^{12}-\left(\frac{\sigma_{0}}{r_{ij}}\right)^{6}
+\frac{1}{4} \right]\Delta(r_{ij}-d_{\mathrm{nat}}) \nonumber
\end{eqnarray}
where $N$ is the number of monomers (residues), $r_{ij}$ is the current distance between
two monomers $i$ and $j$, and $\varepsilon$ is the characteristic attractive energy. The
first term represents rigidity of the backbone; here $d_{0}=3.8${\text{\AA}},
$k_{1}=\varepsilon/{\text{\AA}}^2$, and $k_{2}=100 \varepsilon/{\text{\AA}}^4$. The
second and third terms describe contributions of the native and non-native contacts,
respectively. The contact between two monomers $i$ and $j$ was considered as a native
contact if $|j-i|>1$ and the distance between these monomers in the native state was less
than 7.5{\text{\AA}}. In the second term $\sigma_{ij}=2^{1/6}d_{ij}$, where $d_{ij}$ is
the distance between monomers $i$ and $j$ in the native state, and in the third term
$d_{\mathrm{nat}}=\langle d_{ij}\rangle$, $\sigma_{0}=2^{1/6}d_{\mathrm{nat}}$, and
$\Delta(r_{ij}-d_{\mathrm{nat}})$ is the cutoff function which is equal to 1 for
$r_{ij}<d_{\mathrm{nat}}$ and 0 otherwise. The number of native contacts is equal to 27.

The simulations were performed using a constant-temperature molecular dynamics based on
the Langevin equation
\begin{equation} \label{eq1}%
m\frac{d^{2}\mathbf{r}_{i}}{dt^{2}}+\gamma\frac{d\mathbf{r}_{i}}{dt}=%
-\frac{\partial U}{\partial \mathbf{r}_{i}}+\mathbf{\Phi}_{i}(t)
\end{equation}
where $\mathbf{r}_{i}$ is the radius-vector of $i$ monomer representing a residue, $m$ is
the monomer mass, $U$ is the potential energy of the system, $\mathbf{\Phi}_{i}$ is a
random force from the surroundings (a solvent that is not considered explicitly), and
$\gamma$ is the friction coefficient that introduces viscosity of the surroundings to
balance the random force and dissipation. The random forces have the Gaussian
distribution with zero mean and variance $\langle \Phi_{i}^{j}(t)\Phi_{i'}^{j'}(t+\tau)
\rangle=2\gamma k_{\mathrm{B}}T\delta_{ii'}\delta_{jj'}\delta(\tau)$, where the angular
brackets denote an ensemble average, the upper index at $\Phi$ stands for the vector
component, and $\delta_{kk'}$ and $\delta (\tau)$ are the Kronecker and Dirac delta
functions, respectively. The equation was numerically integrated using the algorithm of
Ref. \onlinecite{BiswasHamann86} with the time step $\Delta t=0.005\tau$ and
$\gamma=3m/\tau$. With the length scale $l=7.5${\text{\AA}} and the attractive energy
$\varepsilon \sim 1{\mathrm{kcal/mol}}$, the characteristic time scale
$\tau=(ml^2/\varepsilon)^{1/2}$ is $\sim 1$ps, where $m$=110 Da.

\begin{figure}[h]\centering%
\resizebox{0.6\linewidth}{!}{ \includegraphics*{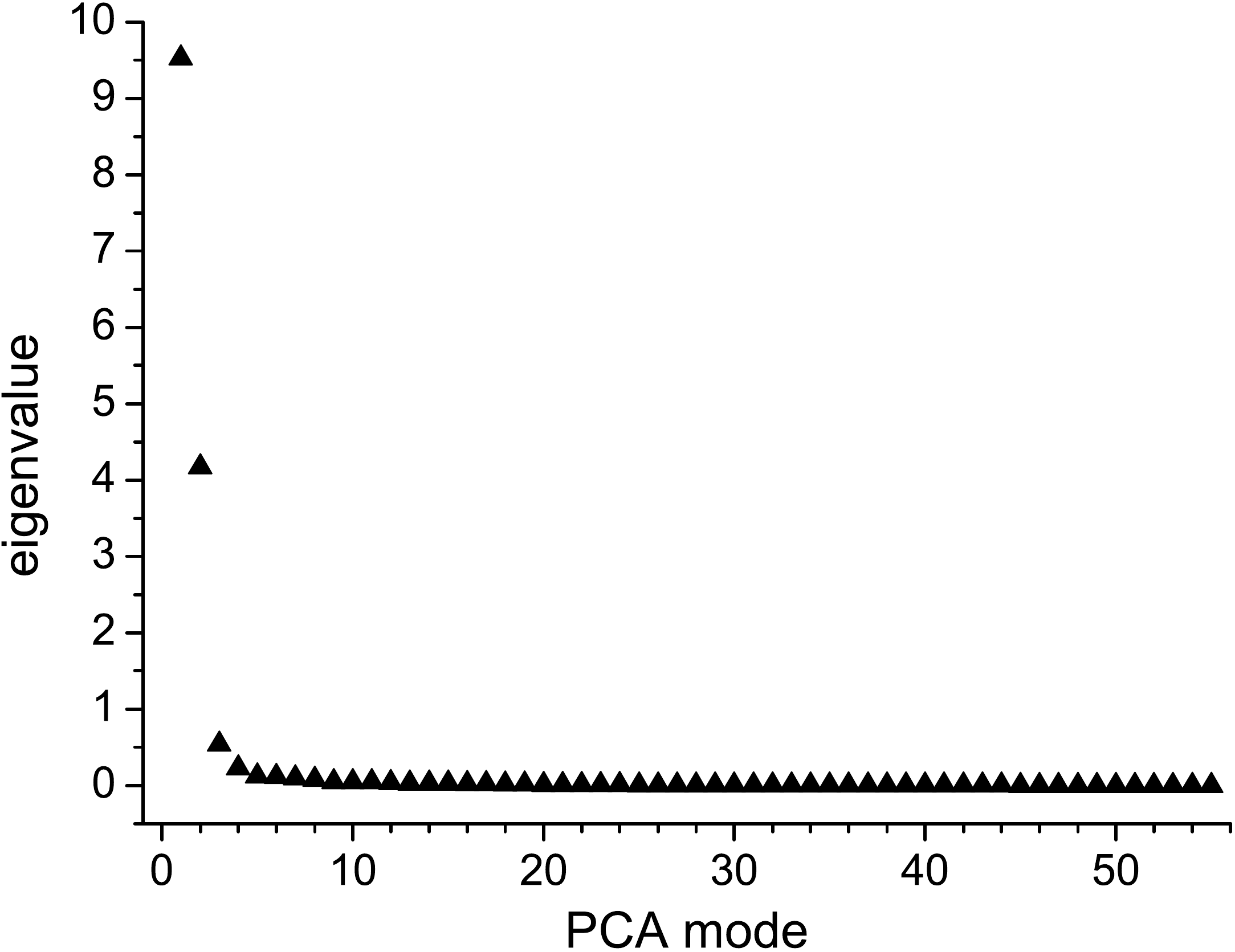}}%
\caption{Spectrum of eigenvalues.} \label{eigenvalues}
\end{figure}

To characterize protein conformations, we employed a set of the bond distances between
the  monomers which are not immediate neighbors along the protein chain; they formed a
55-dimensional conformation space. Using the principal component analysis (PCA)
\cite{Jolliffe02}, this space was reduced to a two-dimensional space of collective
variables $\mathbf{g}=(g_1,g_2)$. The variable $g_1$ was directed along the eigenvector
for the largest eigenvalue, and the variable $g_2$ along the vector calculated as the
linear combination of the rest of the eigenvectors, which contributed with the weights
corresponding to their eigenvalues. Although the second principal component could also be
chosen as $g_2$, because the first two eigenvalues are well separated from the others
(Fig. \ref{eigenvalues}), the present choice makes it possible to take into account the
effects missed by the first component.

The hydrodynamic description of protein folding \cite{ChekmarevPalyanovKarplus08} is
based on the calculation of the transitions in a space of collective variables (the space
$\mathbf{g}$ in the present case). These transitions are organized into transition
probability local flows (fluxes) $\mathbf{j(g)}$. In the case of two variables,
$\mathbf{g}=(g_1, g_2)$, the $g_1$-component of the flux at a point $\mathbf{g}$ is
determined as
\begin{eqnarray} \label{eq2}%
j_{g_1}({\bf g})=[\sum^{{g_1}''-{g_1}'>0}_{{\bf g'},{\bf g''}({\bf g} \subset {\bf
g}^\ast)}n({\bf g''},{\bf g'})  \nonumber \\
-\sum^{{g_1}''-{g_1}'<0}_{{\bf g'},{\bf g''}({\bf g} \subset {\bf g}^\ast)}n({\bf
g''},{\bf g'}) ]/(M\bar{t}_{\mathrm{f}})
\end{eqnarray}
where $M$ is the total number of simulated trajectories, $\bar{t}_{\mathrm{f}}$ is the
mean first-passage time (MFPT), $n({\bf g''},{\bf g'})$ is the number of transitions from
state ${\bf g'}$ to ${\bf g''}$, and ${\bf g} \subset {\bf g}^\ast$ is a symbolic
designation of the condition that the transitions included in the sum have the straight
line connecting points ${\bf g'}$ to ${\bf g''}$, which crosses the line $g_{1}={\mathrm
{const}}$ within the segment of the length of $\Delta g_2$ centered at the point ${\bf
g}$. The $g_2$-component of ${\bf j}(\bf{g})$ is determined in a similar way, except that
one selects the transitions crossing the line ${g_2}={\mathrm {const}}$. The calculations
were performed on a grid with discretization $\Delta g_1=\Delta g_2=0.12$.

There were simulated $1 \times 10^4$ folding trajectories at a temperature equal to
$T=0.17$ (in units $\varepsilon/k_{\mathrm{B}}$, where $k_{\mathrm{B}}$ is the Boltzmann
constant). Each trajectory started at an unfolded state of the protein and was terminated
upon reaching the native state. The unfolded states were prepared by thermalization of
the native conformation at $T=0.5$ for $1 \times 10^4$ time steps; if the number of
native contacts did not exceed 4, the conformation was accepted, if exceeded, the
thermalization was continued. The native state was considered to be reached when the
root-mean square deviation (RMSD) of the protein conformation from the native
conformation was less than 1{\text{\AA}}.

\begin{figure}[h]\centering%
\resizebox{0.6\linewidth}{!}{ \includegraphics*{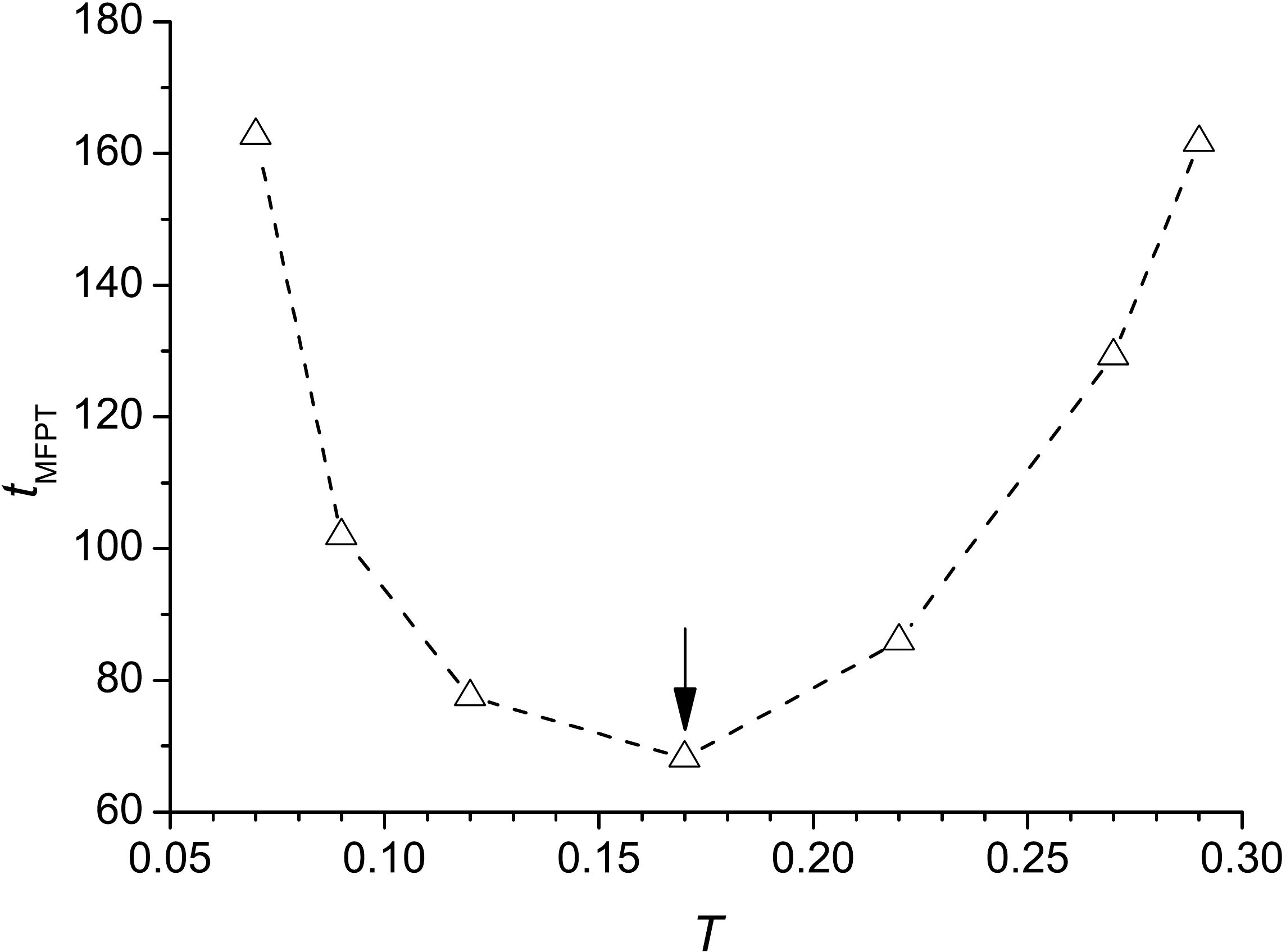}}%
\caption{The mean first-passage time as a function of temperature. The arrow indicates
the temperature ($T=0.17$) at which the study was performed.} \label{mfpt}
\end{figure}

\begin{figure}[h]\centering%
\resizebox{0.6\linewidth}{!}{ \includegraphics*{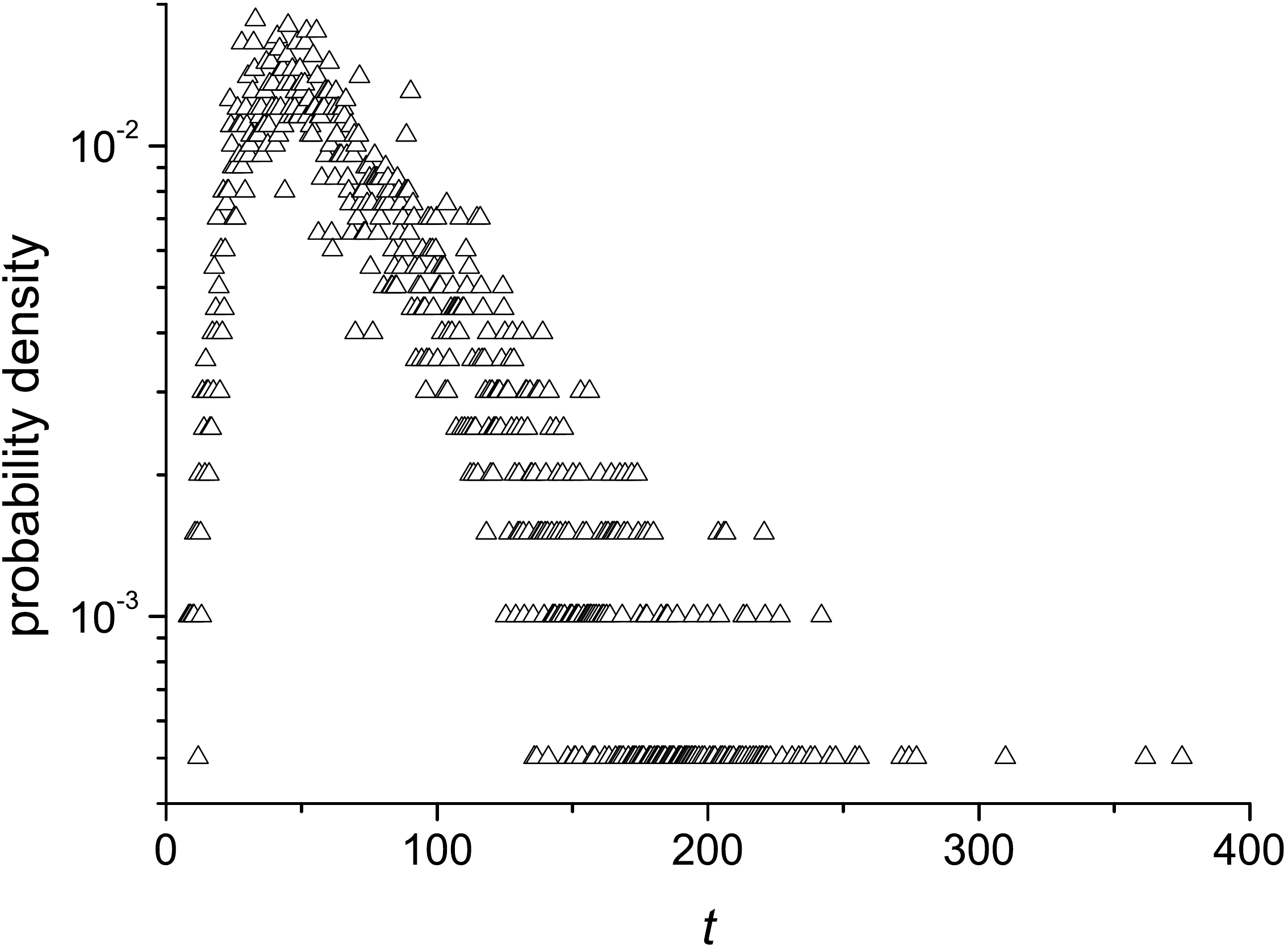}}%
\caption{The distribution of the first-passage times, $T=0.17$.} \label{ftd}
\end{figure}

The simulations have shown that the given protein model provides a sufficiently
representative picture of protein folding, both from the "thermodynamic" and kinetic
viewpoints. The FES calculated as a function of the fraction of native contacts and the
radius of gyration is "L-shaped", which is characteristic of a wide family of proteins in
which an early collapse is observed \cite{SheaBrooks01}, including beta-proteins
\cite{PandeRokhsar99,DinnerLazaridisKarplus99} (Fig. \ref{fes_nat_rg}), the dependence of
the MFPT on the temperature exhibits the well-known U-shape behavior found in experiments
and simulations \cite{OlivebergYanFersht95,Karplus97,ChekmarevKrivovKarplus05} (Fig.
\ref{mfpt}), and the folding kinetics are essentially single-exponential at the "optimal"
folding temperature at which the MFPT has the minimum value
\cite{ChekmarevKrivovKarplus05} (Fig. \ref{ftd}).

\section{Results and Discussion}
\subsection{Free Energy Surface and Folding Flow Field in the PCA Variables}
\begin{figure}[h]\centering%
\resizebox{0.75\linewidth}{!}{ \includegraphics*{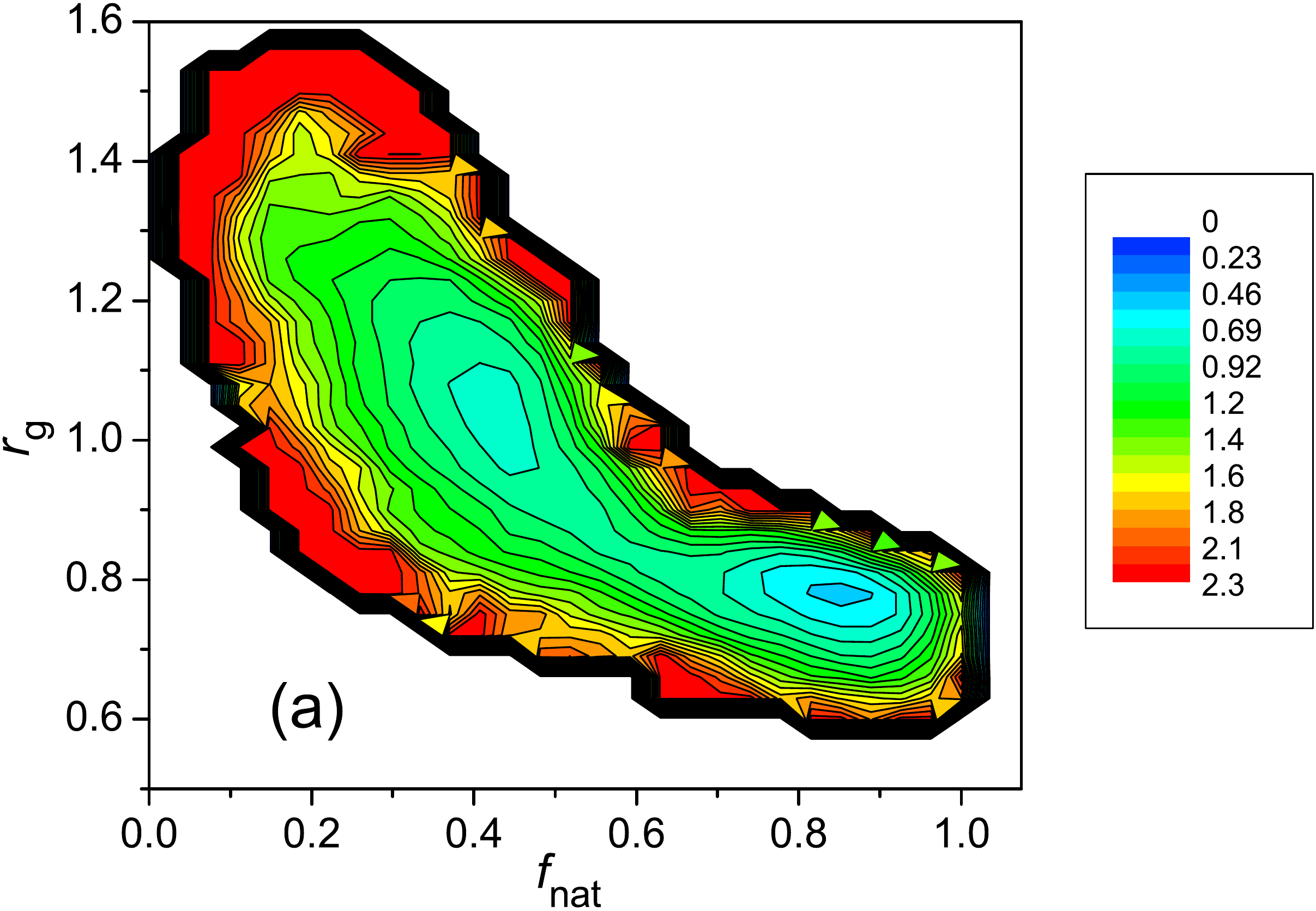}}%
\caption{The free energy surface as a function of the fraction of native contacts and
radius of gyration, $T=0.17$.} \label{fes_nat_rg}
\end{figure}

\begin{figure}\centering%
\resizebox{0.75\linewidth}{!}{ \includegraphics*{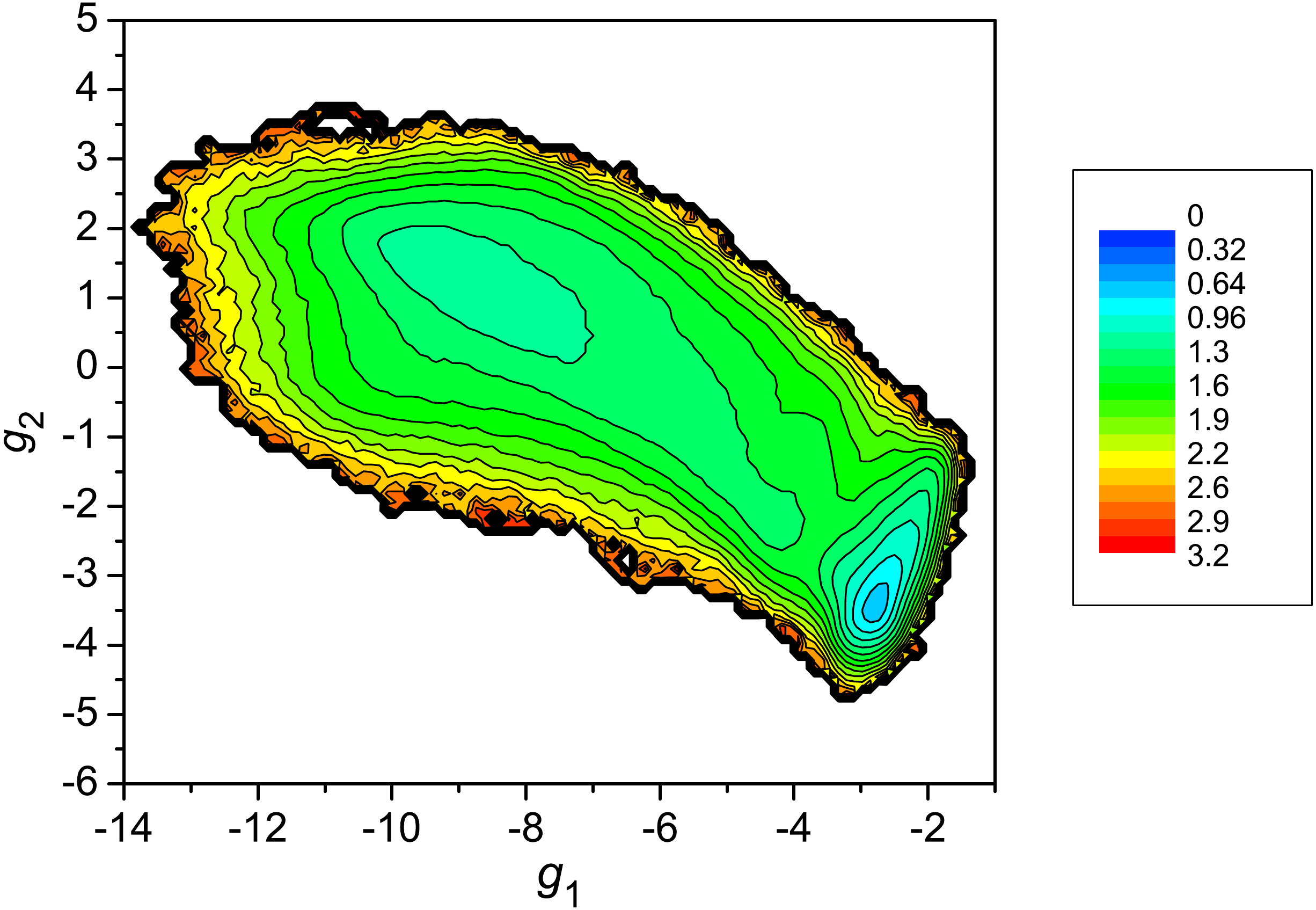}}%
\caption{Free energy surface as a function of collective variables $g_1$ and $g_2$.}
\label{fes}
\end{figure}

The FES as a function of the collective PCA variables $g_1$ and $g_2$ is shown in Fig.
\ref{fes}. It was calculated similar to the above shown FES in "physical" variables (Fig.
\ref{fes_nat_rg}), i.e., using the equation similar to Eq. (\ref{eq0}). The FES has a
characteristic "bean-like" shape typically observed if one employs the PCA coordinates as
the collective variables, e.g., Refs. \onlinecite{KrivovKarplus04} and
\onlinecite{LevyJortnerBecker01}. At the same time, the present FES retains two essential
properties of the FES of Fig. \ref{fes_nat_rg}, in that the global landscape of the
surface has the form of a valley connecting the unfolded states and the native state, and
there exists a free energy barrier separating these states.

\begin{figure}[b]\centering%
\resizebox{0.6\linewidth}{!}{ \includegraphics*{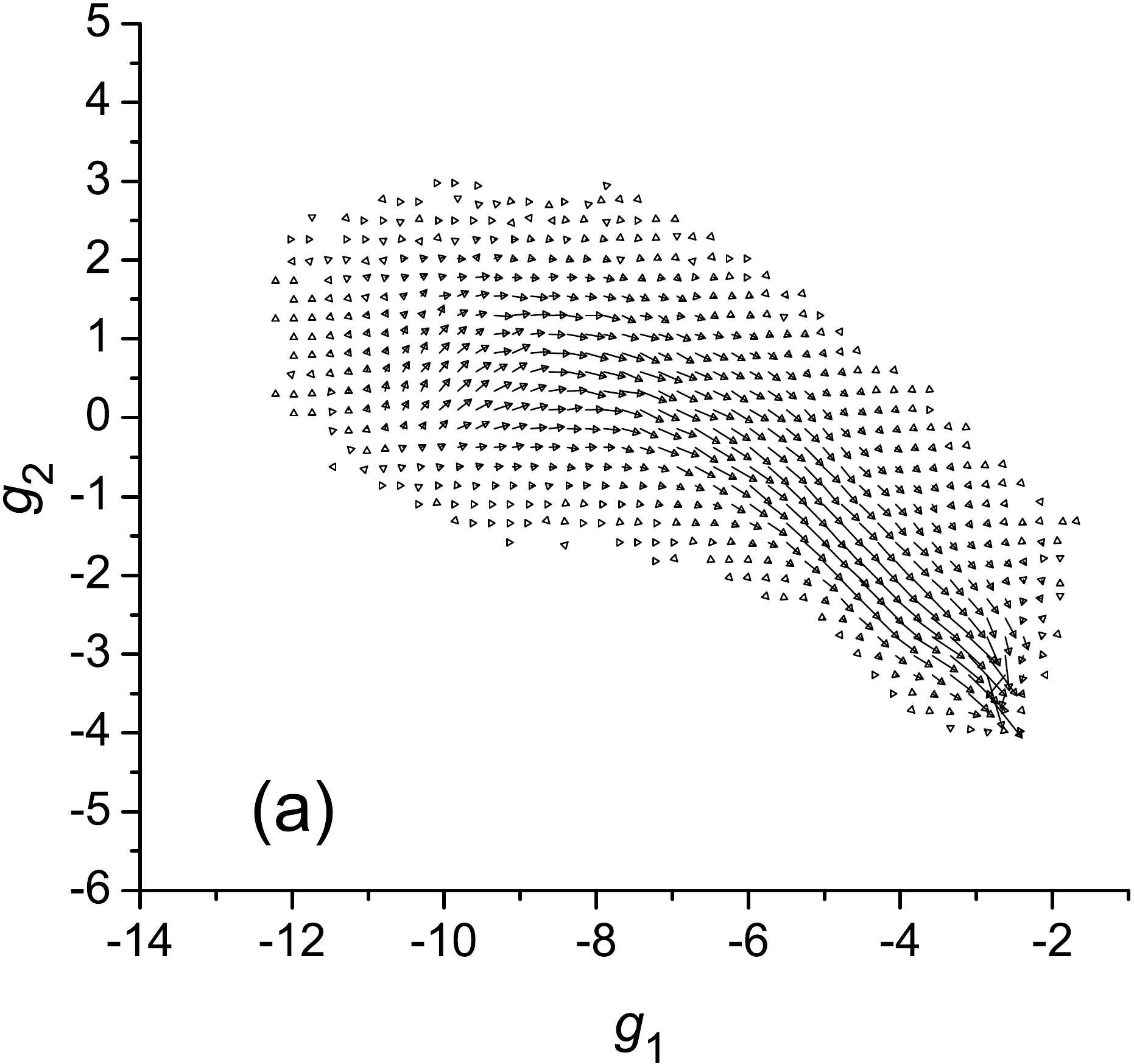}}%
\vfill
\resizebox{0.6\linewidth}{!}{ \includegraphics*{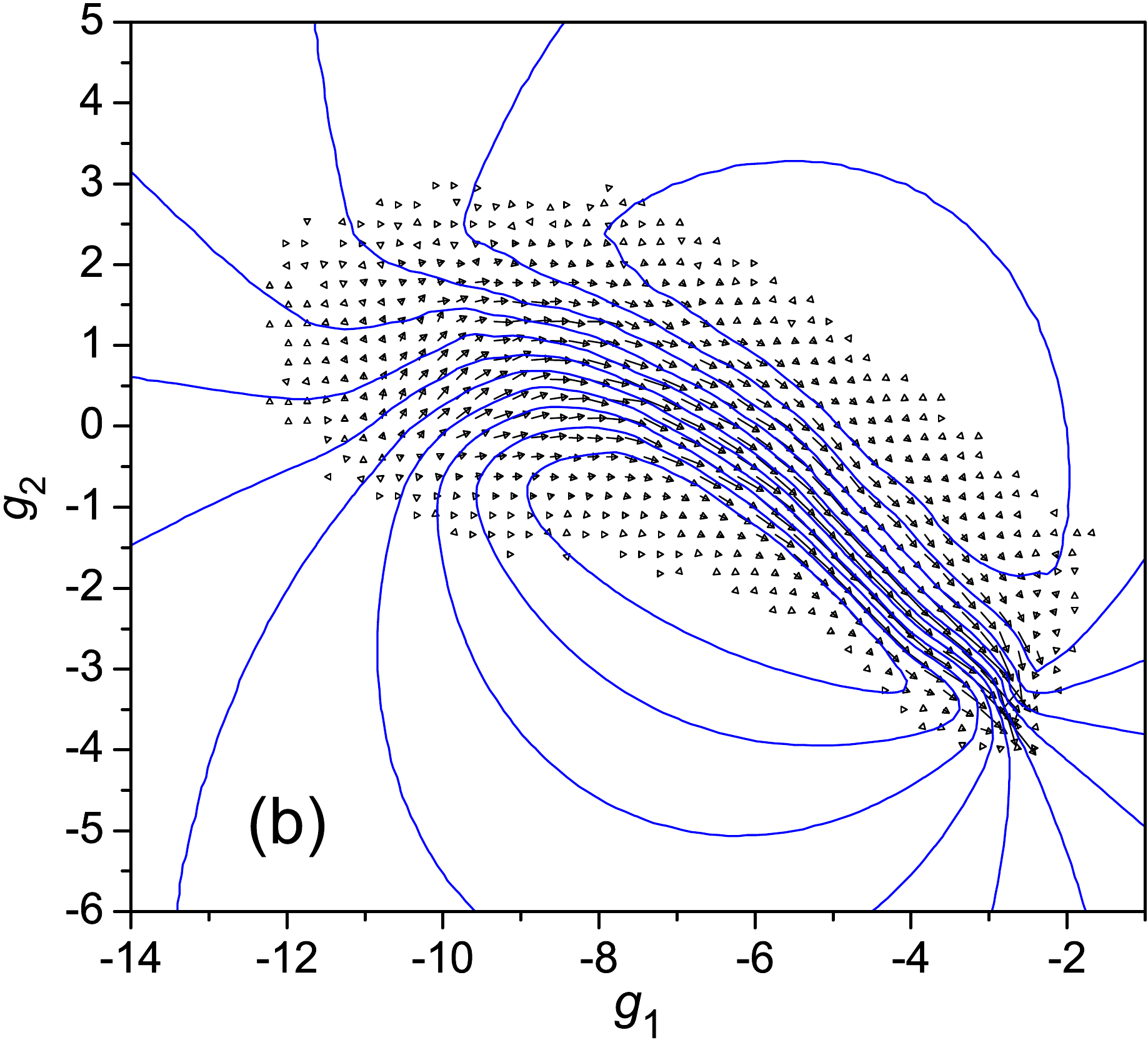}}%
\caption{Vector field of folding fluxes: ($\bf{a}$) the vector field, and ($\bf{b}$) the
streamlines $\Psi(\mathbf{g})=\mathrm{const}$ (blue lines) superimposed on the vector
field. The fraction of the total flow in a stream tube between two neighboring
streamlines is equal to 0.1.} \label{arrows}
\end{figure}

\begin{figure}[h]\centering%
\resizebox{0.6\linewidth}{!}{ \includegraphics*{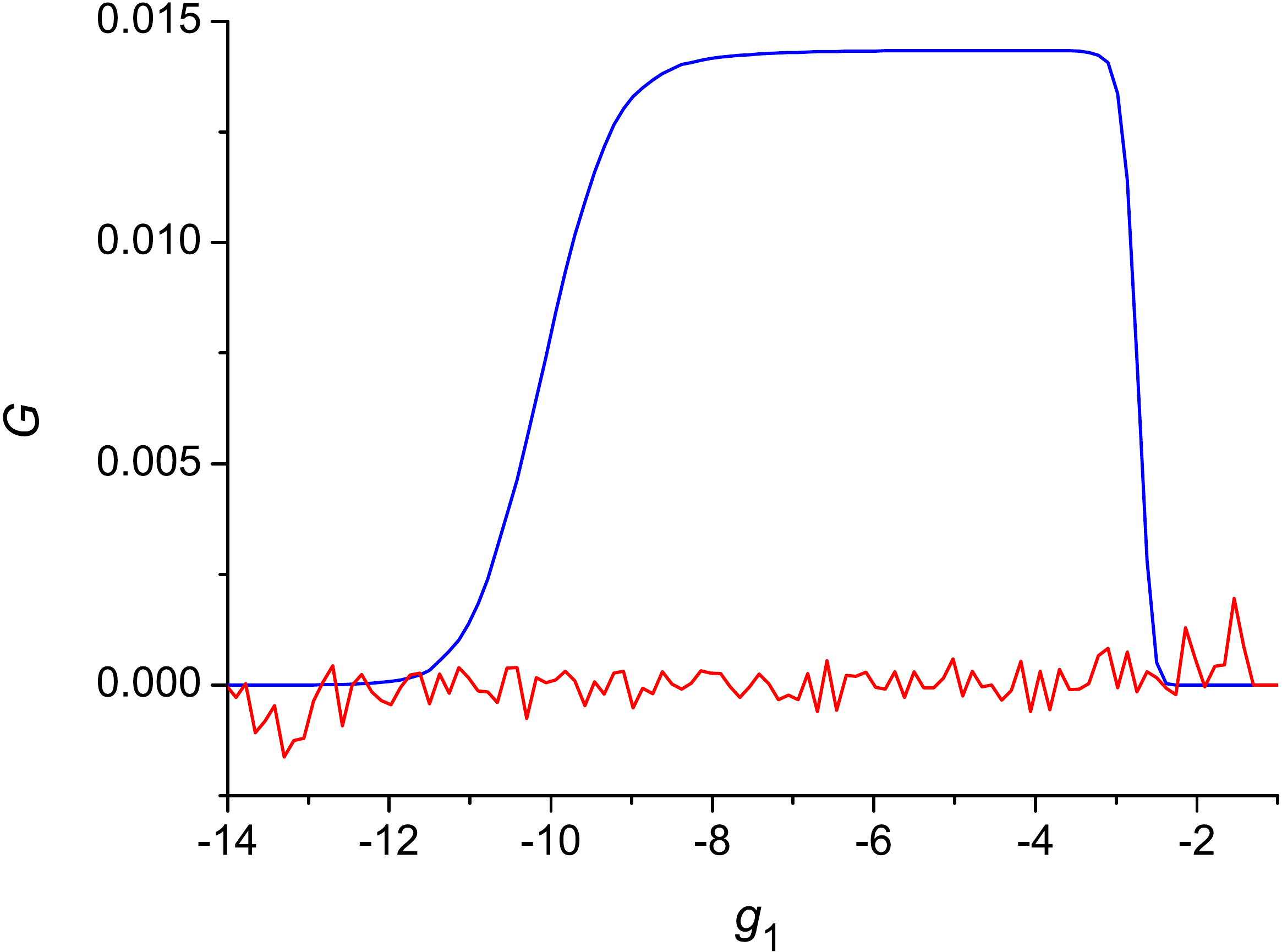}}%
\caption{The total flow from the unfolded states to the native state as a function of the
longitudinal variable $g_1$. The blue curve depicts the total flow as it was obtained in
the simulations. The product of the maximum value of the flow ($G \approx 1.4338 \times
10^{-2}$) by the MFPT ($t_{\mathrm{f}} \approx 69.7368$) is close to 1 with very good
accuracy. The red curve shows the total flow calculated from the fluxes of Fig.
\ref{arrows_from_fes}; the value of the flow is reduced by $1 \times 10^4$ times. }
\label{discharge}
\end{figure}

Figure \ref{arrows}$\bf{a}$ depicts the distribution of $\mathbf{j(g)}$ in the form of
vector field (for illustrative purpose, the lengths of the vectors are equally increased
by factor $3.5 \times 10^2$). As is seen from the comparison of Fig. \ref{arrows}$\bf{a}$
with Fig. \ref{fes}, the flow field lies within the free energy valley that connects the
unfolded states to the native state, and the flow is concentrated at the bottom of the
valley. Integration of the $g_1$-component of $\mathbf{j(g)}$ over $g_2$ in each
cross-section $g_1={\mathrm{const}}$ shows that the total flow from the unfolded states
to the native state, $G(g_1)=\int j_{g_1}(g_1,g_2)dg_2$, is essentially constant in the
region between the source and sink of the flow (Fig. \ref{discharge}).

\begin{figure}[b]\centering%
\resizebox{0.75\linewidth}{!}{ \includegraphics*{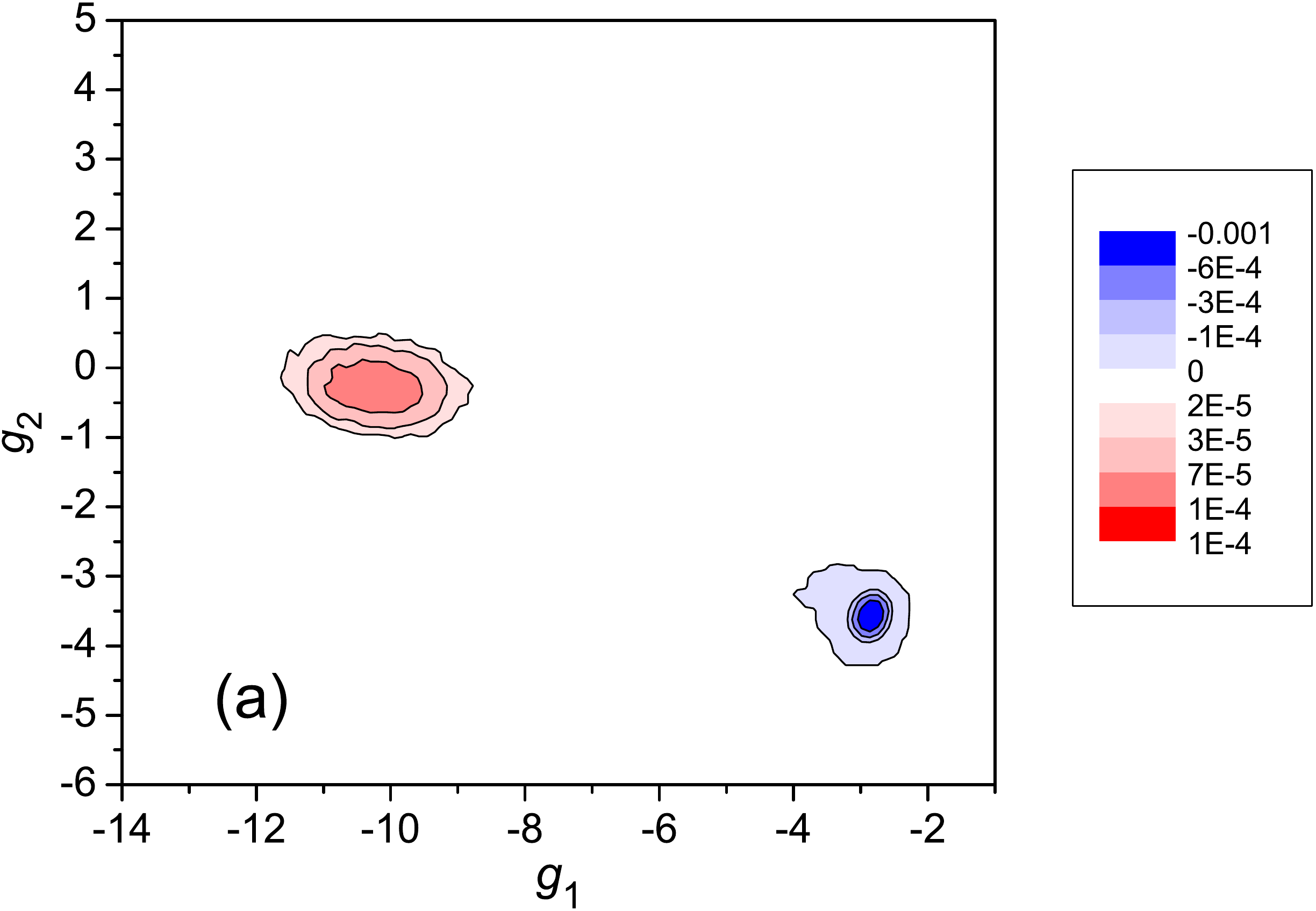}}%
\vfill
\resizebox{0.75\linewidth}{!}{ \includegraphics*{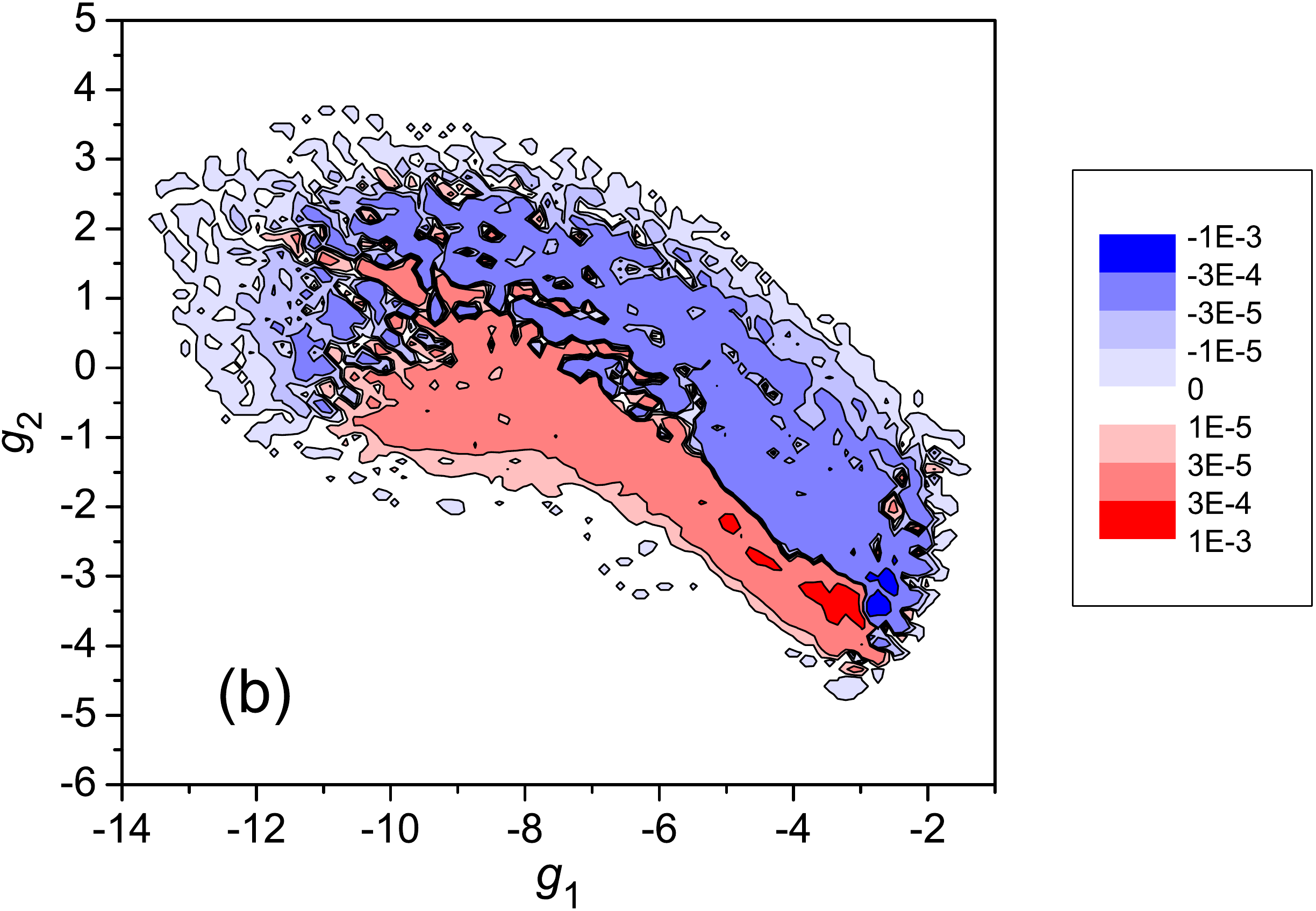}}%
\caption{The distributions of the ($\bf{a}$) divergence and ($\bf{b}$) vorticity of the
folding fluxes.} \label{div_vor}
\end{figure}

Figure \ref{div_vor} shows how the divergence of the folding flow ($q=\partial
j_{g_1}/\partial g_1+\partial j_{g_2}/\partial g_2$, panel {\textbf{a}}) and its
vorticity ($\omega=\partial j_{g_2}/\partial g_1-\partial j_{g_1}/\partial g_2$, panel
{\textbf{b}}) are distributed in the $(g_1,g_2)$ space. Except for two localized regions
at the unfolded and native states, which represent, respectively, the source and sink of
the folding flow, the flow is divergence-free. The vorticity, in contrast, spans the
entire flow field. It arises because the intensity of the flow decreases towards both
sides of the free energy valley (Fig. \ref{arrows}$\bf{a}$); correspondingly, the
vorticity has different signs on the different sides of the valley (see the above
equation for vorticity). This decrease of the flow intensity towards the valley sides is
a natural phenomenon, because the lower the probability to visit some region of the
conformation space (in the present case, a side of the valley), the smaller the flows in
this region; similar nonuniform distributions of the flows have been previously observed
for an $\alpha$-helical hairpin \cite{ChekmarevPalyanovKarplus08} and SH3 domain
\cite{KalginKarplusChekmarev09}. Therefore, the vorticity generated on the periphery of
the folding flow field presents an intrinsic property of the folding dynamics.

\begin{figure}[h]\centering%
\resizebox{0.6\linewidth}{!}{ \includegraphics*{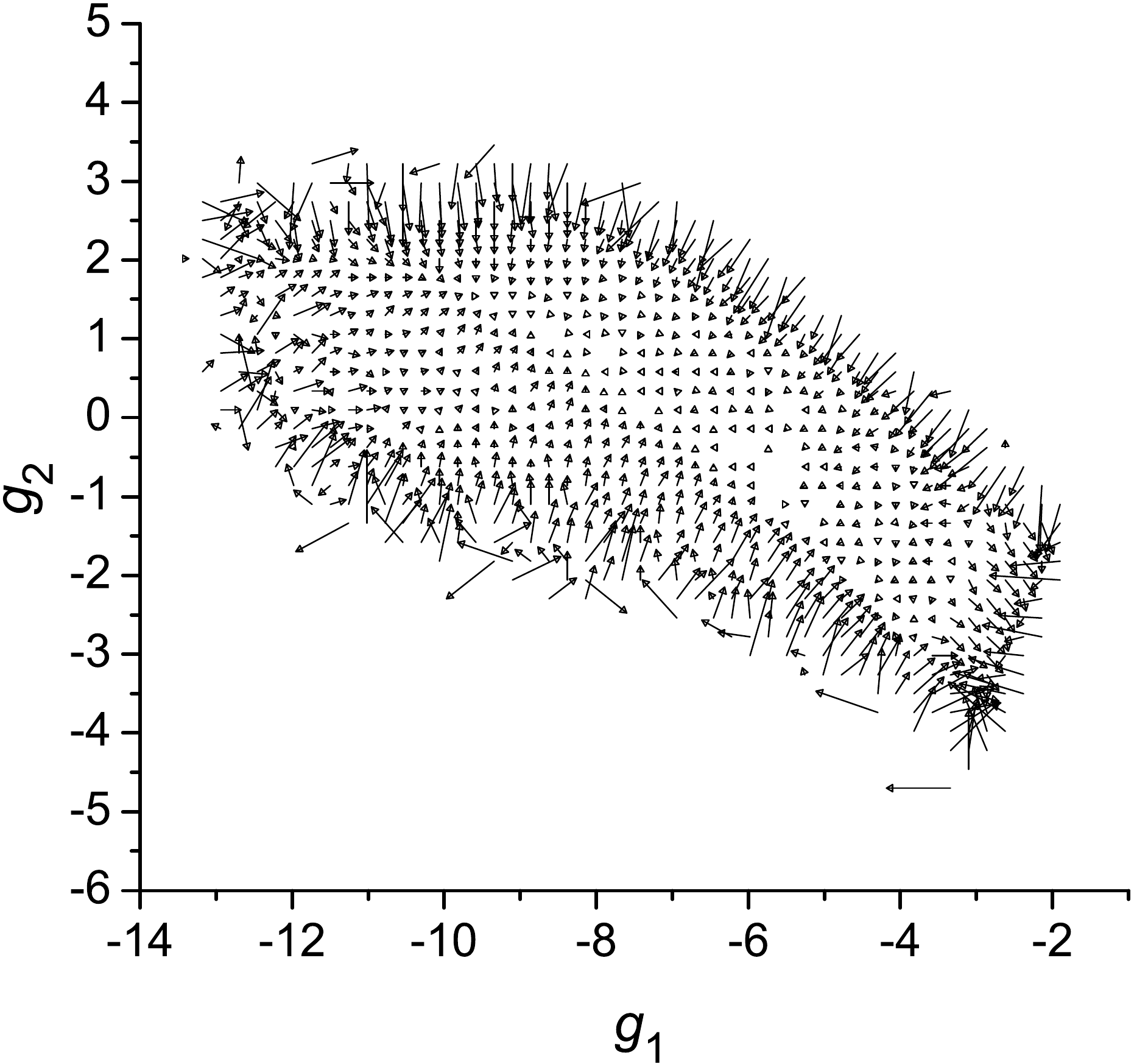}}%
\caption{Vector flow field calculated from the FES of Fig. \ref{fes_nat_rg} with the
fluxes determined as $j_{g_1}=-\partial F(\mathbf{g})/\partial g_1$ and
$j_{g_2}=-\partial F(\mathbf{g})/\partial g_2$, where $F(\mathbf{g})$ is the free energy
function of Fig. \ref{fes_nat_rg}.} \label{arrows_from_fes}
\end{figure}

Comparison of Fig. \ref{arrows}$\bf{a}$ to Fig. \ref{fes} shows that the folding flux
distribution is consistent with the FES landscape in that the overall folding flow
follows the valley of the FES that connects the unfolded states to the native state. At
the same time, the corresponding free energy $F(\mathbf{g})$ does not present a true
potential of folding fluxes, i.e., determining the fluxes as $j_{g_1}=-\partial
F(\mathbf{g})/\partial g_1$ and $j_{g_2}=-\partial F(\mathbf{g})/\partial g_2$ leads to a
vector flow field drastically different from that of Fig. \ref{arrows}$\bf{a}$ (Fig.
\ref{arrows_from_fes}). Calculation of the corresponding total flow shows that the flow
is not constant but fluctuates around zero value (Fig. \ref{discharge}).

\subsection{Potentials for Folding Fluxes}
To determine an actual potential for the folding fluxes, we use the Helmholtz
decomposition theorem \cite{ArfkenWeber95}, according to which any smooth vector field
can be uniquely represented as a sum of two terms: a curl-free field and a
divergence-free field. Then
\begin{equation}\label{eq3}
\mathbf{j}={\mathbf{j}}_{\mathrm{cf}}+{\mathbf{j}}_{\mathrm{df}}
\end{equation}
where ${\mathbf{j}}_{\mathrm{cf}}$ is the curl-free component, and
${\mathbf{j}}_{\mathrm{df}}$ is the divergence-free component, i.e., $\nabla \times
{\mathbf{j}}_{\mathrm{cf}}=0$ and $\nabla \cdot {\mathbf{j}}_{\mathrm{df}}=0$,
respectively. These conditions allow introducing the potentials of the vector fields. In
the case of two dimensions, vectors ${\mathbf{j}}_{\mathrm{cf}}$ and
${\mathbf{j}}_{\mathrm{df}}$ can be written as
\begin{eqnarray}\label{eq3_1}
{\mathbf{j}}_{\mathrm{cf}}=-\frac{\partial \Phi}{\partial g_1}{\mathbf{k}}_{1}%
-\frac{\partial \Phi}{\partial g_2}{\mathbf{k}}_{2}\\ \nonumber
{\mathbf{j}}_{\mathrm{df}}=\frac{\partial \Psi}{\partial g_2}{\mathbf{k}}_{1}%
-\frac{\partial \Psi}{\partial g_1}{\mathbf{k}}_{2}
\end{eqnarray}
where $\Phi=\Phi({\mathbf{g}})$ is the potential for the curl-free component,
$\Psi=\Psi({\mathbf{g}})$ is the potential for the divergence-free component, and
${\mathbf{k}}_{1}$ and ${\mathbf{k}}_{2}$ are the unit vectors for the variables $g_1$
and $g_2$, respectively. The sets of the equipotential lines
$\Phi(\mathbf{g})=\mathrm{const}$ and $\Psi(\mathbf{g})=\mathrm{const}$ are not mutually
orthogonal because $\nabla \Psi \cdot \nabla \Phi=\partial \Psi/\partial g_1 \cdot
\partial \Phi/\partial g_1 + \partial \Psi/\partial g_2 \cdot \partial \Phi/\partial g_2$
is generally nonzero. This is the effect of the flow vorticity: if the flow were both
divergence- and curl-free, the flow flux ${\mathbf{j}}$ could be written in either form
of Eq. (\ref{eq3_1}). Then, the potential functions $\Phi({\mathbf{g}})$ and
$\Psi({\mathbf{g}})$ would satisfy the relations $\partial \Psi/\partial g_2=-\partial
\Phi/\partial g_1$ and $\partial \Psi/\partial g_1=\partial \Phi/\partial g_2$
characteristic of the potential flows (the Cauchy-Riemann conditions)
\cite{LandauLifshitz87}, in which case $\nabla \Psi \cdot \nabla \Phi=0$.

Substituting the given expressions for ${\mathbf{j}}_{\mathrm{cf}}$ and
${\mathbf{j}}_{\mathrm{df}}$ into Eq. (\ref{eq3}) and regrouping the terms, one obtains
\begin{equation}\nonumber
\mathbf{j}=j_{g_1}{\mathbf{k}}_{1}+j_{g_2}{\mathbf{k}}_{2}
\end{equation}
where
\begin{equation}\label{eq4}
j_{g_1}=-\frac{\partial \Phi}{\partial g_1}+\frac{\partial \Psi}{\partial g_2},
\hspace{0.5cm} j_{g_2}=-\frac{\partial \Phi}{\partial g_2}-\frac{\partial \Psi}{\partial
g_1}
\end{equation}

To find functions $\Phi(\mathbf{g})$ and $\Psi(\mathbf{g})$, the functional
\begin{eqnarray} \label{eq5}
Q=\int [\left(j_{g_1}+\frac{\partial \Phi}{\partial g_1}-\frac{\partial
\Psi}{\partial g_2}\right)^2 \nonumber \\%
+\left(j_{g_2}+\frac{\partial \Phi}{\partial g_2}+\frac{\partial \Psi}{\partial
g_1}\right)^2]dg_1dg_2
\end{eqnarray}
was minimized with respect to $\Phi(\mathbf{g})$ and $\Psi(\mathbf{g})$; here $j_{g_1}$
and $j_{g_2}$ are the folding fluxes obtained with the molecular dynamics simulations.
Integration in Eq. (\ref{eq5}) was performed numerically on a grid with the same
discretization as for Eq. (\ref{eq2}). To avoid boundary effects, the region of
integration was extended to $-21.8 \le g_1 \le 7.0$ and $-13.7 \le g_2 \le 12.7$
({\it{cf.}} Fig. \ref{fes}). At the boundaries of the region, no-flux conditions were
imposed, i.e., at the left and right boundaries ($g_1=\mathrm{const}$) it was assumed
$j_{g_1}=0$, and at the lower and upper boundaries ($g_2=\mathrm{const}$) $j_{g_2}=0$.
The minimization was performed using the steepest-descent method with a variable
step-size. Starting with $\Phi(\mathbf{g})=\Psi(\mathbf{g})=0$ as an initial guess for
these functions, in which case $Q \approx 1 \times 10^{-3}$, the process was continued
until $Q$ was as small as $\approx 1 \times 10^{-11}$, i.e., even in the worst case, when
the deviation of the fluxes determined by Eqs. (\ref{eq4}) from the fluxes obtained in
the simulations was concentrated at a single point, it would not exceed $\sim 0.01 \%$.

\begin{figure}\centering%
\resizebox{0.6\linewidth}{!}{ \includegraphics*{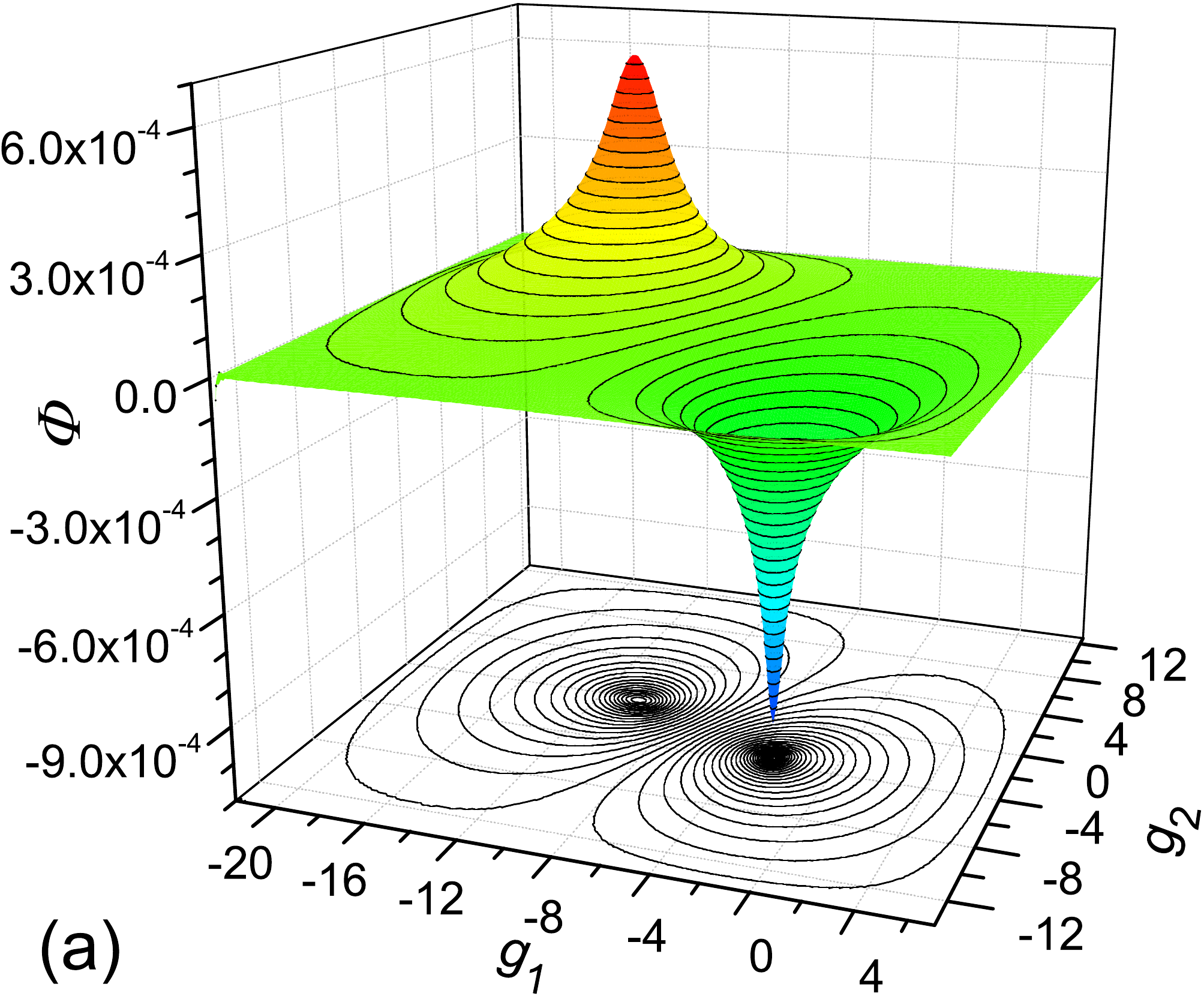}}%
\vfill
\resizebox{0.6\linewidth}{!}{ \includegraphics*{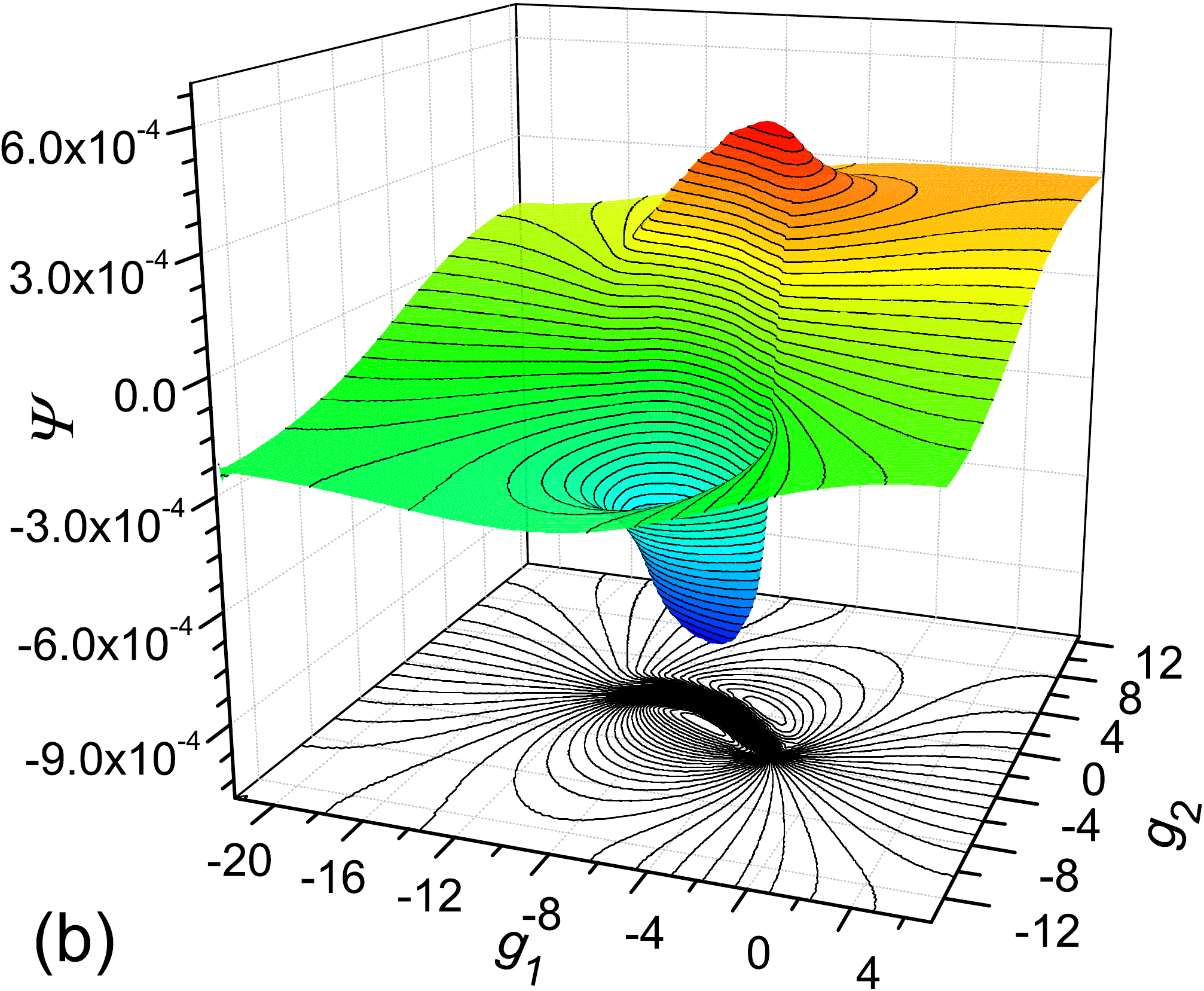}}%
\caption{The functions $\Phi(g_1,g_2)$ (panel $\bf{a}$) and $\Psi(g_1,g_2)$ (panel
$\bf{b}$).} \label{phi_psi}
\end{figure}

The results of the calculations are shown in Fig. \ref{phi_psi}. In agreement with the
Helmholtz decomposition \cite{ArfkenWeber95}, the functions $\Phi(\mathbf{g})$ and
$\Psi(\mathbf{g})$ are characteristically different in that the former accounts for the
source and sink of the flows, and the latter for the vorticity effects. More
specifically, as can be seen from comparison of Eq. (\ref{eq4}) with Fig. \ref{phi_psi},
the function $\Phi(\mathbf{g})$ determines the intensity of the flow in the vicinity of
the source and sink, and the function $\Psi(\mathbf{g})$ determines it in the region
between the source and sink, providing the canalization of the flow within the free
energy valley that connects the unfolded states to the native state. Thus, both
functions, $\Phi(\mathbf{g})$ and $\Psi(\mathbf{g})$, are equally important for a correct
description of the folding dynamics and should be considered as the intrinsic components
of the complex potential that determines the driving force.

\subsection{Interpretation of the Potentials}
In hydrodynamic terms \cite{LandauLifshitz87}, the functions $\Phi(\mathbf{g})$ and
$\Psi(\mathbf{g})$ can be associated, respectively, with the scalar potential of the flow
and the vector potential, or more specifically, with the component of the vector
potential that is orthogonal to the two-dimensional plane under consideration. The latter
plays a role of the stream function, for which the equation
$\Psi(\mathbf{g})=\mathrm{const}$ determines the streamline of the flow, i.e., the line
that is tangent to the local directions of the $\mathbf{j(g)}$ vectors (Fig.
\ref{arrows}$\bf{b}$). The difference between the stream functions for two streamlines
determines the fraction of the total flow in the stream tube between the streamlines (see
also Refs. \onlinecite{ChekmarevPalyanovKarplus08} and
\onlinecite{KalginKarplusChekmarev09}). Similar to a steady potential flow of an inviscid
fluid in a two-dimensional space \cite{LandauLifshitz87}, the functions
$\Phi(\mathbf{g})$ and $\Psi(\mathbf{g})$ can also be written as a complex function
$\Theta({\mathbf{g}})=\Phi(\mathbf{g})+i\Psi(\mathbf{g})$, where $i$ is the imaginary
unit; then ${\mathbf{j}}({\mathbf{g}})=j_{g_1}+ij_{g_2}=-\nabla_{c}
\Theta({\mathbf{g}})$, where $\nabla_{c}=\partial/\partial g_1+i\partial/\partial g_2$ is
the gradient operator in the complex number space $(g_1,ig_2)$. However, in contrast to
the potential flow of inviscid fluid, the function $\Theta({\mathbf{g}})$ is not {\it
analytic} function, i.e., is not a function of complex variable $g_1+ig_2$, because the
Cauchy-Riemann conditions, which represent the necessary and sufficient condition for a
function to be analytic and require the flow to be divergence- and curl-free
\cite{ArfkenWeber95}, are not satisfied (Fig. \ref{div_vor}). Therefore, the methods of
the theory of analytic functions of complex variable, which are successfully used in
hydrodynamics \cite{LandauLifshitz87}, are not applicable here.

Another interpretation of the potentials $\Phi(\mathbf{g})$ and $\Psi(\mathbf{g})$ stems
from the kinetic theory \cite{Kampen81}. According to Eq. (\ref{eq4}), the velocity of
motion is proportional to the forces produced by these potentials, i.e., the local flows
are assumed to be {\it drift} flows. In other words, although the inertia of the monomers
was present in the Langevin equation we used for the simulations, the potentials
$\Phi(\mathbf{g})$ and $\Psi(\mathbf{g})$ determined from the resulted fluxes $j_{g_1}$
and $j_{g_2}$ account for an {\it overdamped} motion. The approximation of the overdamped
motion, which neglects the inertia term in the Langevin equation (\ref{eq1}), is rather
common to characterize the protein folding dynamics \cite{RohrdanzZhengClementi13}. The
corresponding kinetic equation is the Smoluchowski equation $\partial p/\partial t+\nabla
\cdot \mathbf{J}=0$, where $p(\mathbf{g},t)$ is the probability density, $\nabla$ is the
gradient operator in the coordinate space $\mathbf{g}$, and $\mathbf{J}$ is the
probability current, which can be written as $\mathbf{J}=-D(\mathbf{g})\nabla
p+D(\mathbf{g})/(k_{\mathrm B}T){\mathbf F}(\mathbf{g})p$, where $-D(\mathbf{g})\nabla p$
is the diffusion flux, $D(\mathbf{g})/(k_{\mathrm B}T){\mathbf F}(\mathbf{g})p$ is the
drift flux, $D(\mathbf{g})$ is the diffusion tensor, and ${\mathbf F}(\mathbf{g})$ is the
driving force. If the probability current were zero, i.e., detailed balance existed, the
equality $\mathbf{J}=0$ would give the {\it equilibrium} (Boltzmann) distribution
$p(\mathbf{g}) \sim \exp[-G(\mathbf{g})]/k_{\mathrm B}T$, where $G(\mathbf{g})$ is the
free energy that exerts the driving force ${\mathbf F}(\mathbf{g})=-\nabla
G(\mathbf{g})$. This case would correspond to a curl-free drift flow, with the potential
of the driving force being determined by a single function in the form of the free energy
$G(\mathbf{g})$. However, if the probability current is nonzero, as in the present case,
i.e., when the steady flow from the unfolded state to the native state exists, the
stationary solution is determined by the condition $\nabla \cdot \mathbf{J}=0$, so that
$\mathbf{J}$ can have a curl component \cite{Kampen81}. Such flow is {\it
non-equilibrium} and is characterized by "irreversible circulation" or "cyclic balance",
which can be considered as a measure of deviation from detailed balance
\cite{TomitaTomita74,Graham77,EyinkLebowitzSpohn96}. In our case the circulating flow is
represented by the flux vector ${\mathbf{j}}_{\mathrm{df}}$ and, according to Eq.
(\ref{eq3_1}), is generated by the potential $\Psi(\mathbf{g})$, with the factor
$D(\mathbf{g})/(k_{\mathrm B}T)$ being included into the potential. In general case, the
driving force can be written as $\mathbf{F}=-\nabla \Phi+\nabla \times \mathbf{A}$, where
$\Phi$ and $\mathbf{A}$ are the scalar and vector potentials, respectively (in the case
of two dimensions, only the component of $\mathbf{A}$ that is orthogonal to the plane is
involved). The potentials of this type have recently been used to study the dynamics of
Brownian particles in corrugated channels
\cite{MartensStraubeSchmidSchimansky-GeierHaanggi13,AiHeLiZhong13}.

\begin{figure}\centering%
\resizebox{0.6\linewidth}{!}{ \includegraphics*{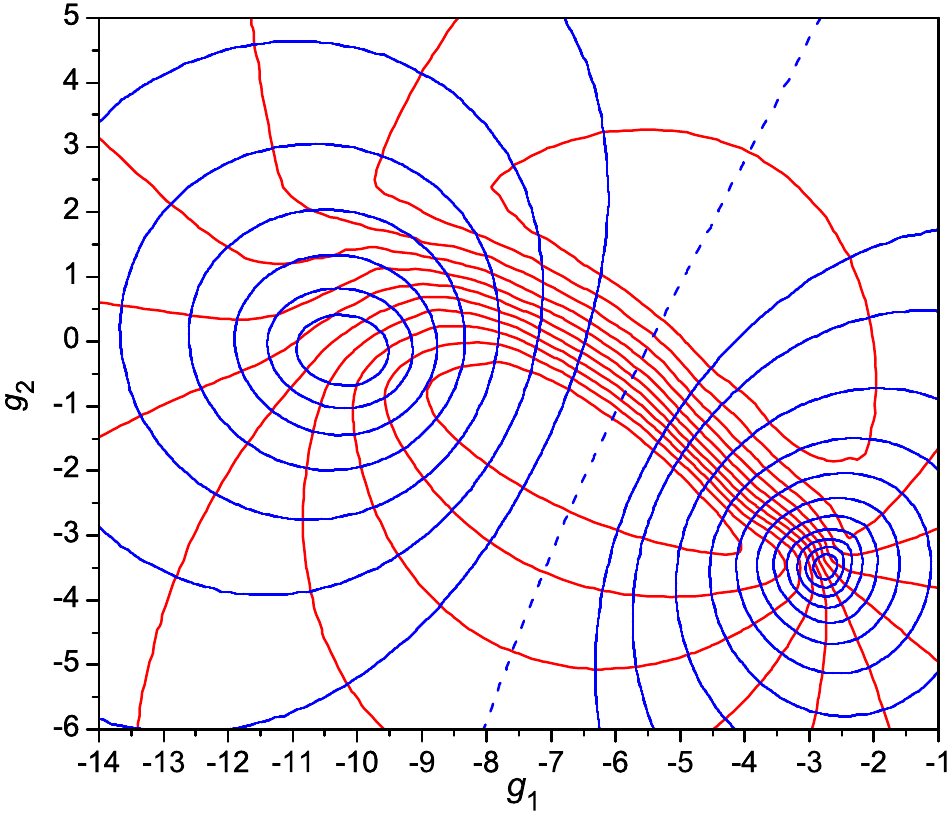}}%
\caption{The sets of the equipotential lines $\Phi(g_1,g_2)={\mathrm{const}}$ (blue
color) and $\Psi(g_1,g_2)={\mathrm{const}}$ (red color) corresponding to Fig.
\ref{phi_psi}. The dashed line is for $\Phi(g_1,g_2)=0$.} \label{phi_psi_2d}
\end{figure}

As an anonymous reviewer of the manuscript remarked, the resulting picture of the folding
process in terms of the $\Psi$ and $\Phi$ potentials is very similar to the picture
obtained for the reaction dynamics in the TPT \cite{E_Vanden-Eijnden06}, although the
approaches are apparently different. In the latter, the committor probability functions
are calculated for the transition paths from the reactant to the product, and the spatial
distribution of the probability currents of the reaction paths (fluxes) between these
states is determined \cite{E_Vanden-Eijnden10}. The currents are normal to the
isocommittor surfaces and organized in the form of the reaction tubes that connect the
reactant and product wells, so that the boundaries of the tubes represent the flow lines
(streamlines) of the currents. Then the equipotential lines
$\Psi(\mathbf{g})=\mathrm{const}$ and $\Phi(\mathbf{g})=\mathrm{const}$ (Fig.
\ref{phi_psi_2d}) can be associated with the flow lines and the isocommittor surfaces,
respectively. In particular, the line $\Phi(\mathbf{g})=0$, shown by the dashed curve,
which separates the source and sink regions of the flow, can play a role of the
isocommittor surface of probability of 1/2. Very interesting from physical and
methodological viewpoints, this analogy, however, does not probably extend beyond
qualitative resemblance because, in contrast to the flow lines and the isocommittor
surfaces, the sets of the equipotential lines $\Phi(\mathbf{g})=\mathrm{const}$ and
$\Psi(\mathbf{g})=\mathrm{const}$ are not mutually orthogonal (as has been previously
indicated). As is seen from Fig. \ref{phi_psi_2d}, they possess the orthogonality
property only approximately.

\section{Conclusions}
In summary, we have shown that the potential of the driving force for protein folding in
a two-dimensional space of collective variables has two components. One component
accounts for the source and sink of the folding flow, representing, respectively, the
unfolded and native states of the protein, and the other accounts for the vorticity
generated at the periphery of the flow field and provides the canalization of the flow
within the free energy valley that connects the unfolded and native states. Since the
Helmholtz decomposition theorem \cite{ArfkenWeber95} guarantees that any vector field can
be uniquely represented by a sum of divergence-free and curl-free fields, the present
approach is equally applicable to more complex folding dynamics that are characterized by
multiple reaction channels and formation local vortices of the folding flows, as, for
example, for folding of SH3 domain \cite{KalginKarplusChekmarev09,KalginChekmarev11}. In
this case, the functions $\Phi(\mathbf{g})$ and $\Psi(\mathbf{g})$ will be not as regular
as in Fig. \ref{phi_psi} because they should take into account this additional complexity
of the folding flow, but the general pattern of these functions is expected to be
preserved because of the presence of the source and sink of the folding flow and its
vorticity inherently generated at the periphery of the flow field. The requirement of
smoothness of the vector field, which is necessary for the Helmholtz decomposition
\cite{ArfkenWeber95}, currently restricts the possibilities of analysis of folding of
proteins of practical interest on atomic level of resolution, however, for small
proteins, such, e.g., as the beta3s miniprotein, the calculation of smooth vector fields
is quite feasible \cite{KalginCaflischChekmarevKarplus13}. Also note that the application
of the present approach is not limited to the case of the folding reaction. Similar
complex potentials can be expected for other systems that involve multiple reaction
pathways, e.g., for nanoclusters, in which the transition between the structures of
alternative morphology is characterized by a multi-funnel energy landscape
\cite{Wales03}.

\section{Acknowledgments}
I would like to thank an anonymous reviewer for the remark on the similarity between the
hydrodynamic and TPT pictures of the reaction.

\newpage

\begin{references}
\bibitem{BryngelsonWolynes87} J. D. Bryngelson and P. G. Wolynes, Proc. Natl. Acad. Sci.
USA \textbf{84}, 7524 (1987).

\bibitem{ShakhnovichFarztdinovGutinKarplus91} E. Shakhnovich, G. Farztdinov, A. M. Gutin,
and M. Karplus, Phys. Rev. Lett. \textbf{67}, 1665 (1991).

\bibitem{LeopoldMontalOnuchic92} P. E. Leopold, M. Montal, and J. N. Onuchic,
Proc. Natl. Acad. Sci. USA \textbf {89}, 8721 (1992).

\bibitem{BryngelsonOnuchicSocciWolynes95} J. D. Bryngelson, J. N. Onuchic,
N. D. Socci, and P. G. Wolynes, Proteins: Structure, Function, and Bioinformatics
\textbf{21}, 167 (1995).

\bibitem{DillChan97} K. A. Dill and H. S. Chan, Nat. Struct, Biol. \textbf{4},
10 (1997).

\bibitem{Shakhnovich97} E. I. Shakhnovich, Curr. Opin. Struct. Biol. \textbf{7}, 29
(1997).

\bibitem{OnuchicLuthey_SchultenWolynes97} J. N. Onuchic, Z.
Luthey-Schulten, P. G. Wolynes, Annu. Rev. Phys. Chem. \textbf{48}, 545 (1997).

\bibitem{DobsonSaliKarplus98} C. M. Dobson, A. \u{S}ali, and M. Karplus,
Angew. Chem. Int. Ed. \textbf{37}, 869 (1998).

\bibitem{DSSDK00} A. R. Dinner, A. \u{S}ali, L. J. Smith,
C. M. Dobson, and M. Karplus, Trends Biochem. Sci. \textbf{25}, 331 (2000).

\bibitem{SheaBrooks01}  J.-E. Shea and C. L. Brooks III,
Annu. Rev. Phys. Chem. \textbf {52}, 499 (2001).

\bibitem{SchulerEverettLipmanEaton02} B. Schuler, E. A. Lipman, and W. A. Eaton,
Nature \textbf{419}, 743 (2002).

\bibitem{Gruebele2002} M. Gruebele, Curr. Opinion Struct. Biol. \textbf{12}, 161 (2002).

\bibitem{MelloBarrick04} C. C. Mello and D. Barrick, Proc. Natl. Acad. Sci. USA
\textbf {101}, 14102 (2004).

\bibitem{Karplus11} M. Karplus, Nat. Chem. Biol. {\textbf{7}}, 401 (2011).

\bibitem{DillOzkanShellWeikl08} K. A. Dill, S. B. Ozkan, M. S. Shell, and T. R. Weikl,
Annu. Rev. Biophys. \textbf{37}, 289–316 (2008).

\bibitem{DillMacCallum12} K. A. Dill and J. L. MacCallum, Science \textbf{338},
1042 (2012).

\bibitem{FinkelsteinPtitsyn02} A. V. Finkelstein and O. Ptitsyn, {\it Protein Physics:
A Course of Lectures} (Academic Press: London, 2002).

\bibitem{BerezhkovskiiSzabo05} A. Berezhkovskii and A. Szabo, J. Chem. Phys. \textbf{122},
014503 (2005).

\bibitem{RheePande05} Y. M. Rhee and V. S. Pande, J. Phys. Chem. B \textbf{109},
6780 (2005).

\bibitem{BestHummer2006} R. B. Best and G. Hummer, Proc. Natl. Acad. Sci. U.S.A.
\textbf{102}, 6732 (2005).

\bibitem{KrivovKarplus06} S. V. Krivov and M. Karplus, J. Phys. Chem. B \textbf{110},
12689 (2006).

\bibitem{BeckDaggett07} D. A. C. Beck and V. Daggett, Biophys. J. \textbf{93}, 3382 (2007).

\bibitem{ChoderaPande11} J. D. Chodera and V. S. Pande, Phys. Rev. Lett. \textbf{107},
098102 (2011).

\bibitem{PetersBolhuisMullenShea13} B. Peters, P. G. Bolhuis, R. G. Mullen,
and J.-E. Shea, J. Chem. Phys. \textbf{138}, 054106 (2013).

\bibitem{E_Vanden-Eijnden10} W. E and E. Vanden-Eijnden, Annu. Rev. Phys. Chem. \textbf{61},
391 (2010).

\bibitem{BoczkoBrooks95} E. M. Boczko and C. L. Brooks III, Science \textbf{269},
393 (1995).

\bibitem{SocciOnuchicWolynes98}  N. D. Socci, J. N. Onuchic, and P. G. Wolynes,
Proteins: Structure, Function, and Genetics \textbf {32}, 136 (1998).

\bibitem{ChekmarevKrivovKarplus05} S. F. Chekmarev, S. V. Krivov, and M. Karplus,
J. Phys. Chem. B {\textbf{109}}, 5312 (2005).

\bibitem{ChekmarevPalyanovKarplus08} S. F. Chekmarev, A. Yu. Palyanov, and M. Karplus,
Phys. Rev. Lett. {\textbf{100}}, 018107 (2008).

\bibitem{Schutte99} C. Sch\"{u}tte, Habilitation thesis, Department of Mathematics and Computer
Science, Freie Universitat Berlin, 1999.

\bibitem{SwopePiteraSuits04} W. C. Swope, J. W. Pitera, and F. Suits, J. Phys. Chem. B
{\textbf{108}}, 6571 (2004).

\bibitem{WeberPande11} J. K. Weber and V. S. Pande, J. Chem. Theory Comput. {\textbf{7}},
3405 (2011).

\bibitem{BeckerKarplus97} O. M. Becker and M. Karplus, J. Chem. Phys. {\textbf{106}},
1495 (2007).

\bibitem{KrivovKarplus04} S. V. Krivov and M. Karplus, Proc. Natl. Acad. Sci. U.S.A.
\textbf{101}, 14766 (2004).

\bibitem{Wales03} D. J. Wales, {\it Energy Landscapes: Applications to Clusters,
Biomolecules and Glasses} (Cambridge University Press: Cambridge, 2003).

\bibitem{RaoCaflisch04} F. Rao and A. Caflisch, J. Mol. Biol. {\textbf{342}}, 299 (2004).

\bibitem{Noe-Weikl} F. No\'{e}, C. Sch\"{u}tte, E. Vanden-Eijnden, L. Reich, and T. R.
Weikl, Curr. Opin. Struct. Biol. {\textbf{106}}, 19011 (2009).

\bibitem{BowmanVoelzPande11} G. R. Bowman, V. A. Voelz, and V. S. Pande,
Proc. Natl. Acad. Sci. USA {\textbf{21}}, 4 (2011).

\bibitem{KalginCaflischChekmarevKarplus13} I. V. Kalgin, A. Caflisch, S. F. Chekmarev,
and M. Karplus, J. Phys. Chem. B {\textbf{117}}, 6092 (2013).

\bibitem{LandauLifshitz87} L. D. Landau and E. M. Lifshitz, {\it Fluid Mechanics}
(Pergamon: New York, 1987).

\bibitem{KalginKarplusChekmarev09} I. V. Kalgin, M. Karplus, and S. F. Chekmarev,
J. Phys. Chem. B {\textbf{113}}, 12759 (2009).

\bibitem{KalginChekmarev11} I. V. Kalgin and S. F. Chekmarev, Phys. Rev. E
{\textbf{83}}, 011920 (2011).

\bibitem{E_Vanden-Eijnden06} W. E and E. Vanden-Eijnden, J. Stat. Phys. \textbf{123},
503 (2006).

\bibitem{ArfkenWeber95} G. B. Arfken and H. J. Weber, {\it Mathematical Methods for
Physicists}, 4th ed. (Academic Press: San Diego, 1995).

\bibitem{2evq} N. H. Andersen, K. A. Olsen, R. M. Fesinmeyer, X. Tan, F. M. Hudson,
L. A. Eidenschink, and S. R. Farazi, J. Am. Chem. Soc. \textbf{128}, 6101 (2006).

\bibitem{Go83} N. G\={o}, Annu. Rev. Biophys. Bioeng. \textbf {12},
183 (1983).

\bibitem{HoangCieplak00} T. X. Hoang and M. Cieplak, J. Chem.
Phys. {\textbf{112}}, 6851 (2000).

\bibitem{BiswasHamann86} R. Biswas and D. R. Hamann, Phys. Rev. B {\textbf{34}},
895 (1986).

\bibitem{Jolliffe02} I. T. Jolliffe, {\it Principal Component Analysis};
2nd ed. (Springer: New York, 2002).

\bibitem{PandeRokhsar99} V. S. Pande and D. S. Rokhsar, Proc. Natl. Acad. Sci.
USA \textbf{94}, 9062 (1999).

\bibitem{DinnerLazaridisKarplus99} A. R. Dinner, T. Lazaridis and M. Karplus,
Proc. Natl. Acad. Sci. USA \textbf{96}, 9068 (1999).

\bibitem{OlivebergYanFersht95} M. Oliveberg, Y.-J. Yan, and A. R. Fersht,
Proc. Natl. Acad. Sci. U.S.A. {\textbf {92}}, 8926 (1995).

\bibitem{Karplus97} M. Karplus, Fold. Des. {\textbf{2}}, 569 (1997).

\bibitem{LevyJortnerBecker01} Y. Levy, J. Jortner, and O. M. Becker,
Proc. Natl. Acad. Sci. U.S.A. \textbf {98}, 2188 (2001).

\bibitem{Kampen81} N. G. van Kampen, {\it Stochastic Processes in Physics and Chemistry}
(North-Holland: Amsterdam, 1981), p. 294.

\bibitem{RohrdanzZhengClementi13} M. A. Rohrdanz, W. Zheng, and C. Clementi,
Annu. Rev. Phys. Chem. \textbf {64}, 295 (2013).

\bibitem{TomitaTomita74} K. Tomita and H. Tomita, Prog. Theor. Phys. \textbf {51},
1731 (1974).

\bibitem{Graham77} R. Graham, Z. Phys. B \textbf {26}, 397 (1977).

\bibitem{EyinkLebowitzSpohn96} G. L. Eyink, J. L. Lebowitz, and H. Spohn, J. Stat. Phys.
\textbf {83}, 385 (1996).

\bibitem{MartensStraubeSchmidSchimansky-GeierHaanggi13} S. Martens, A. V. Straube,
G. Schmid, L. Schimansky-Geier, and P. Ha\"{a}nggi, Phys. Rev. Lett. \textbf {110},
010601 (2013).

\bibitem{AiHeLiZhong13} B.-Q. Ai, Y.-F. He, F.-G. Li, and W.-R. Zhong, J. Chem. Phys.
\textbf {138}, 154107 (2013).

\end{references}
\end{document}